\documentclass[%
  conference,%
  10pt,%
]{IEEEtran}

\usepackage{cite}
\usepackage{amsmath,amssymb,amsfonts}
\allowdisplaybreaks

\usepackage{booktabs}
\usepackage{url}

\usepackage{paralist}

\usepackage[UKenglish]{babel}

\usepackage{graphicx}
\graphicspath{{./images/}}
\usepackage{placeins}
\usepackage{subcaption}
\usepackage{enumitem}
\usepackage{balance}
\usepackage{booktabs}  %
\usepackage{colortbl}

\usepackage{algorithm}%
\usepackage{algpseudocode}%
\usepackage{dirtree}

\usepackage{inconsolata}

\usepackage[hyperindex,breaklinks]{hyperref}
\hypersetup{
   breaklinks=true,   
   pdfusetitle=true,  
}

\usepackage[%
  group-minimum-digits=4,%
  list-final-separator={, and },%
  add-integer-zero=false,%
  free-standing-units,%
  binary-units,%
  detect-weight=true,%
  detect-inline-weight=math,%
]{siunitx}

\usepackage{float}
\newfloat{algorithm}{t}{lop}

\usepackage{microtype}
\usepackage[utf8]{inputenc}

\usepackage[colorinlistoftodos,prependcaption,textsize=tiny
]{todonotes}

\usepackage{xspace}

\usepackage[capitalise]{cleveref}

\crefformat{footnote}{#2\footnotemark[#1]#3}

\usepackage{tabularx}

\newcommand{\multilinecell}[1]{%
    \begin{tabular}{@{}l@{}}#1\end{tabular}
}

\newcommand{\multilinecellr}[1]{%
    \begin{tabular}{@{}r@{}}#1\end{tabular}
}

\usepackage{csquotes}
\usepackage{csvsimple}
\usepackage{listings}
\usepackage{xcolor}

\definecolor{codegreen}{rgb}{0,0.6,0}
\definecolor{codegray}{rgb}{0.5,0.5,0.5}
\definecolor{codepurple}{rgb}{0.58,0,0.82}
\definecolor{backcolour}{rgb}{0.95,0.95,0.92}

\lstdefinestyle{mystyle}{
  backgroundcolor=\color{backcolour},
  commentstyle=\color{codegreen},
  keywordstyle=\color{magenta},
  numberstyle=\tiny\color{codegray},
  stringstyle=\color{codepurple},
  basicstyle=\ttfamily\footnotesize,
  breakatwhitespace=false,
  breaklines=true,
  breakindent=10pt,
  captionpos=b,
  keepspaces=true,
  numbers=left,
  numbersep=5pt,
  showspaces=false,
  showstringspaces=false,
  showtabs=false,
  tabsize=2,
  columns=fullflexible
}

\newcommand{\linebreakand}{%
  \end{@IEEEauthorhalign}
  \hfill\mbox{}\par
  \mbox{}\hfill\begin{@IEEEauthorhalign}
}

\begin{document}

\title{FlaPy: Mining Flaky Python Tests at Scale}

\author{%
 \IEEEauthorblockN{Martin Gruber}%
 \IEEEauthorblockA{%
  \textit{BMW Group, University of Passau}\\%
  Munich, Germany\\%
  martin.gr.gruber@bmw.de%
 }%
 \and%
 \IEEEauthorblockN{Gordon Fraser}%
 \IEEEauthorblockA{%
  \textit{University of Passau}\\%
  Passau, Germany\\%
  gordon.fraser@uni-passau.de%
 }
}

\maketitle

\begin{abstract}

Flaky tests obstruct software development, and studying and proposing mitigations against them has therefore become an important focus of software engineering research. 
To conduct sound investigations on test flakiness,
it is crucial to have large, diverse, and unbiased datasets of flaky tests.
A common method to build such datasets is by rerunning the test suites of selected projects multiple times and checking for tests that produce different outcomes.
While using this technique on a single project is mostly straightforward, applying it to a large and diverse set of projects raises several implementation challenges such as
(1)~isolating the test executions,
(2)~supporting multiple build mechanisms,
(3)~achieving feasible run times on large datasets, and
(4)~analyzing and presenting the test outcomes.
To address these challenges we introduce FlaPy, a framework for researchers to mine flaky tests in a given or automatically sampled set of Python projects by rerunning their test suites.
FlaPy isolates the test executions using containerization and fresh execution environments to simulate real-world CI conditions and to achieve accurate results.
By supporting multiple dependency installation strategies, it promotes diversity among the studied projects.
FlaPy supports parallelizing the test executions using SLURM, making it feasible to scan thousands of projects for test flakiness.
Finally, FlaPy analyzes the test outcomes to determine which tests are flaky and depicts the results in a concise table.
A demo video of FlaPy is available at \mbox{\url{https://youtu.be/ejy-be-FvDY}}

\end{abstract}


\section{Introduction}%
\label{sec:introduction}

Flaky tests are tests that behave non-deterministically, i.e., they can both pass and fail on the same software version.
Test flakiness is a serious issue since it slows down development by requiring manual failure analysis and delaying test feedback when rerunning tests, and because it causes developers to lose trust in testing~\cite{gruber2022survey}.
To create an understanding of test flakiness, researchers have conducted empirical studies by mining and investigating flaky tests in open-source Java projects~\cite{luo2014empirical,bell2018deflaker,lam2019idflakies,shi2019ifixflakies}.
However, the community is lacking mining tools for other programming languages that include features important for research, such as parallelization of test executions.

To promote and enable research on flakiness in other languages, we developed \textit{FlaPy}, a tool for mining flaky Python tests at scale by repeatedly executing the test suites of a given set of Python projects.
Its core features are:
\begin{itemize}
    \item Automatically sampling projects from PyPI~\cite{pypi}
    \item Isolating test executions for realistic and accurate results
    \item Supporting different dependency installation strategies to promote diversity within the set of sample projects
    \item Parallelizing test executions via SLURM~\cite{yoo2003slurm}
    \item Analyzing and presenting results in a concise format
\end{itemize}

\section{Using FlaPy}%
\label{sec:usingFlaPy}

FlaPy is an easy to use tool that allows researchers to identify flaky tests within a given set of Python projects and their test suites.
FlaPy itself is also written in Python and provides a command-line interface to execute tests and to parse the outcomes.
Provided with a set of Python projects as input (\cref{sec:input}), it sets up an isolated executed environment and runs the projects' test suites a specified number of times (\cref{sec:run_tests_locally,sec:run_tests_slurm}).
FlaPy analyzes the resulting test outcomes and presents them in a concise table (\cref{sec:analyzing_results}).

\subsection{Installation}%
\label{sec:installation}

FlaPy uses Docker~\cite{docker} to isolate the test executions and to provide a swift installation experience.
To automate the creation of containers, it comes with supporting scripts which are located in the FlaPy repository and can be cloned via
\begin{lstlisting}[language=bash,backgroundcolor=\color{white},numbers=none]
git clone https://github.com/se2p/flapy
\end{lstlisting}

FlaPy's main entry point is the script \texttt{flapy.sh}.
The FlaPy Docker image will be pulled automatically on first usage.
Note that FlaPy requires Docker to be installed and executable without root privileges.
Apart from that it has no requirements.

\subsection{Input}%
\label{sec:input}

\begin{table*}[ht]
    \centering
    \caption{Input-CSV for our running example.}
    \label{tab:flapyInputCsv}

\resizebox{\textwidth}{!}{
    \begin{tabular}{|l|l|l|l|l|l|}
    \toprule
     PROJECT\_NAME           & PROJECT\_URL                             & PROJECT\_HASH & PYPI\_TAG & FUNCS\_TO\_TRACE & TESTS\_TO\_RUN  \\
    \midrule
      avwx-engine            & https://github.com/avwx-rest/avwx-engine & ed1d3e        &           &                  &                 \\
      avwx-engine            & https://github.com/avwx-rest/avwx-engine & ed1d3e        &           &                  &                 \\
      avwx-engine            & https://github.com/avwx-rest/avwx-engine & ed1d3e        &           &                  &                 \\
      avwx-engine            & https://github.com/avwx-rest/avwx-engine & ed1d3e        &           &                  &                 \\
          jgutils            & https://github.com/jerodg/jgutils        & 6c8ed6        &           &                  &                 \\
          jgutils            & https://github.com/jerodg/jgutils        & 6c8ed6        &           &                  &                 \\
          jgutils            & https://github.com/jerodg/jgutils        & 6c8ed6        &           &                  &                 \\
          jgutils            & https://github.com/jerodg/jgutils        & 6c8ed6        &           &                  &                 \\
    flapy\_example           & ./minimal\_flaky\_python\_tests          &               &           &                  & test\_flaky.py  \\
    flapy\_example           & ./minimal\_flaky\_python\_tests          &               &           &                  & test\_flaky.py  \\
    flapy\_example           & ./minimal\_flaky\_python\_tests          &               &           &                  & test\_flaky.py  \\
    flapy\_example           & ./minimal\_flaky\_python\_tests          &               &           &                  & test\_flaky.py  \\
    \bottomrule
    \end{tabular}

}
\end{table*}

After installing FlaPy, we can define our study subjects, i.e., the set of Python projects whose test suites we want to execute.
As a running example throughout this paper, we use the following projects, in which we previously found flakiness:
\begin{itemize}
    \item \textbf{avwx-engine}: a global aviation weather fetching and parsing engine.
    \item \textbf{jgutils}: a Python utility module.
    \item \textbf{flapy\_example}: an example project created by the authors of this paper, which is located in the FlaPy repository under \texttt{minimal\_flaky\_python\_tests/}
\end{itemize}

For each project, we have to provide a name and a Git-URL from which its sources can be cloned.
Alternatively, we can also provide a path from which the sources will be copied.
Optionally we can specify a Git revision or tag to be checked out before the test execution,
a PyPI~\cite{pypi} version used for installing the project itself to the test environment to retrieve its dependencies instead of searching for dependencies in its source code,
the functions we want to trace (see the documentation for more information),
as well as a subset of tests we want to execute instead of the entire test suite.
This information must be provided via a CSV file, which we call the \mbox{\textit{input-csv}}.
A possible input-csv for our running example is depicted in~\cref{tab:flapyInputCsv} and can also be found in the FlaPy repository under the name \texttt{flapy\_input\_example.csv}.
FlaPy can also generate input-csv files automatically by sampling projects from PyPI~\cite{pypi}. To do so, use \texttt{./flapy.sh sample}.

We refer to one row in the input-csv as an \textit{iteration}.
As~\cref{tab:flapyInputCsv} shows, in our example we do not use one, but four iterations per project.
Splitting the test executions like this has several advantages, particularly when executing the tests on a SLURM cluster (\cref{sec:run_tests_slurm}):
Since each row in the input-csv results in one SLURM job, fine-grained chunking helps to avoid timeouts (many clusters have maximum runtimes per job, e.g., 24h) and allows for more parallelization and better scheduling, speeding up the process.
Furthermore,
since each SLURM job is potentially scheduled on a different machine,
it helps to detect infrastructure flakiness~\cite{gruber2021empirical}, i.e., global issues that lead to bulked test failures.

\subsection{Running Tests Locally}%
\label{sec:run_tests_locally}

To run the test suites of our study subjects, we use the \texttt{run} command on \texttt{flapy.sh}, which offers the following interface:
\begin{lstlisting}[language=bash,backgroundcolor=\color{white},numbers=none]
./flapy.sh run [OPTIONS...] INPUT_CSV NUM_RUNS
\end{lstlisting}

It has two mandatory arguments, namely, the path to the input-csv file (\texttt{INPUT\_CSV}), and the number of times the test suites should be executed (\texttt{NUM\_RUNS}).
Optionally, we can tell FlaPy to conduct additional test executions in random order (\texttt{--plus-random-runs}), allowing it to also detect order-dependent flaky tests, or to write the results to a specific directory (\texttt{--out-dir OUT\_DIR}).
Suppose in our running example we want the tests to be executed five times both in same and in random order and the results to be written to \texttt{example\_results} (folder will be created by a helper script):

\begin{lstlisting}[language=bash,backgroundcolor=\color{white},numbers=none]
./flapy.sh run --out-dir example_results \
  --plus-random-runs flapy_input_example.csv 5
\end{lstlisting}

This causes FlaPy to sequentially iterate over the input-csv and start one container for each iteration (row).
\cref{alg:test_execution} describes the setup and test execution taking place inside the container (entry point \texttt{ITERATION}).
Isolating the test execution is especially important when executing unknown projects, whose code might have unwanted or even malicious side-effects (such as creating cronjobs or deleting files).
Furthermore, test case isolation promotes reproducibility and comparability to real-world CI systems, which also often use Docker.

\algrenewcommand\algorithmicindent{1.0em}%

\makeatletter
\renewcommand{\ALG@beginalgorithmic}{\small}
\makeatother

\everymath{\it}\everydisplay{\it}

\begin{algorithm}
    \caption{Test execution core procedure.}\label{alg:test_execution}

    \begin{algorithmic}[1]
        \Function{iteration}{$proj\_name$, $proj\_url$, $proj\_hash$,\par
            $pypi\_tag$, $tests\_to\_run$, $num\_runs$, $plus\_rand$\par\noindent}
            \State \Call{clone\_or\_copy\_sources}{$proj\_url$}
            \State \Call{checkout\_revision}{$proj\_hash$}
            \State \Call{measure\_lines\_of\_code\_and\_disk\_usage}{\null}
            \For{$i=0$; $i<num\_runs$, $i++$}
                \State \Call{Run}{$proj\_name$, $pypi\_tag$, $tests\_to\_run$, $same$}
            \EndFor
            \If{$plus\_rand$}
                \For{$i=0$; $i<num\_runs$, $i++$}
                    \State \Call{Run}{$proj\_name$, $pypi\_tag$, $tests\_to\_run$, $random$}
                \EndFor
            \EndIf
            \State \Call{archive\_results}{\null}
        \EndFunction
        \\
        \Function{Run}{$proj\_name$, $pypi\_tag$, $tests\_to\_run$, $order$}
            \State \Call{copy\_project\_sources}{\null}
            \State \Call{create\_virtual\_environment}{\null}
            \If{$pypi\_tag$ == \textsc{null}}
                \State \Call{search\_and\_install\_dependencies}{\null}
            \Else
                \State \Call{install\_project}{$proj\_name$, $pypi\_tag$}
            \EndIf
            \State \Call{pytest}{$tests\_to\_run$, $order$}
            \State \Call{delete\_copy\_and\_virtualenv}{\null}
        \EndFunction
    \end{algorithmic}
\end{algorithm}

\everymath{}\everydisplay{}

\begin{figure}
\begin{lstlisting}
example_results                     # OUT_DIR
`-- flapy-results_20221123_161945   # results-dir
    |-- !flapy.run
    |   |-- flapy_run.yaml
    |   `-- input.csv
    |
    |-- avwx-engine_20221123_161945_2       # iteration-dir
    |   |-- flapy-iteration-result.yaml
    |   |-- loc.csv
    |   `-- results.tar.xz
    |       | -- .../avwx-engine_output0.xml
    |       | -- .../avwx-engine_coverage0.xml
    |       | -- .../avwx-engine_output1.xml
    |       ...
    |
    |-- avwx-engine_20221123_161945_3
    |-- avwx-engine_20221123_161945_4
    |-- avwx-engine_20221123_161945_5
    |-- flapy_example_20221123_161945_10
    |-- flapy_example_20221123_161945_11
    |-- flapy_example_20221123_161945_12
    |-- flapy_example_20221123_161945_13
    |-- jgutils_20221123_161945_6
    |-- jgutils_20221123_161945_7
    |-- jgutils_20221123_161945_8
    `-- jgutils_20221123_161945_9
\end{lstlisting}
    \caption{Output structure for running example.}%
    \label{fig:output}
\end{figure}

One execution of \texttt{"./flapy.sh run"} creates one new folder within the \texttt{OUT\_DIR} (\texttt{example\_results/}), which we refer to as a \textit{results-directory}.
Results-directories have the following naming schema: \enquote*{flapy-results\_DATE\_TIME}, where DATE and TIME describe the point in time when the execution started.
The results-directory created by our example execution is \texttt{flapy-results\_20221123\_161945}. %
Its structure is depicted in ~\cref{fig:output}.
Within a results-directory, FlaPy creates one sub-directory for each row in the input-csv, which we refer to as an \textit{iteration-directory}.
Each iteration-directory is named as follows: \enquote*{ProjectName\_DATE\_TIME\_RowNum}, where RowNum is the line number of the iteration within the input-csv.
This helps to differentiate iterations of the same project within the same results-directory.
One such iteration-directory in our example is \texttt{avwx-engine\_20221123\_161945\_2}.

Each iteration-directory contains the following files:
\begin{itemize}
    \item \texttt{flapy-iteration-result.yaml}: Input- and meta-information about this iteration.
    \item \texttt{loc.csv}: The lines of code of this project per programming language, measured via CLOC~\cite{adanial_cloc}. (The raw CLOC sqlite database can be found in \texttt{results.tar.xz})
    \item \texttt{results.tar.xz}: An archive containing the test results (junit-xml and coverage-xml files)
\end{itemize}
\cref{sec:analyzing_results} discusses how to use FlaPy's results-parser to generate an overview of the outcomes of all test executions.

\subsection{Running Tests on SLURM}%
\label{sec:run_tests_slurm}

SLURM~\cite{yoo2003slurm} is a popular open-source cluster management and job scheduling system, which can be used to execute multiple jobs in parallel across different machines.
When executing FlaPy on a SLURM cluster, we highly recommend the usage of Podman instead of Docker.
Podman~\cite{podman} is a Docker alternative that provides the same interface, but allows rootless container executions and parallel non-blocking pulling of images, which happens when multiple FlaPy jobs run in parallel on the same machine.
The use of Podman instead of Docker can be configured in \texttt{setup\_docker\_command.sh}, which is sourced before any Docker commands are invoked, or by setting the \texttt{FLAPY\_DOCKER\_COMMAND\_SETUP\_SCRIPT} environment variable. %
This can also be used to configure other cluster-specific setup steps, for example, (un)setting or changing global variables such as \texttt{HOME} or \texttt{TMP}.

To conduct the test executions on a SLURM cluster, we add the option \texttt{--run-on cluster} and specify a subset of nodes on which the job execution shall be scheduled via \texttt{--constraint}:
\begin{minipage}{\linewidth}
\begin{lstlisting}[language=bash,backgroundcolor=\color{white},numbers=none]
./flapy.sh run --out-dir example_results \
  --plus-random-runs \
  --run-on cluster --constraint CONSTRAINT \
  flapy_input_example.csv 5
\end{lstlisting}
\end{minipage}
Make sure that \texttt{OUT\_DIR} (\texttt{example\_results}) is a shared storage that is accessible from all nodes.

\subsection{Analyzing Results}%
\label{sec:analyzing_results}

With the \texttt{parse} command, FlaPy offers a convenient way to analyze the test outcomes generated by the \texttt{run} command.
To create an overview showing which tests have been executed and how often they yielded a certain verdict, we can use the following command for our running example:
\begin{lstlisting}[language=bash,backgroundcolor=\color{white},numbers=none]
./flapy.sh parse ResultsDirCollection \
  --path example_results \
  get_tests_overview _df \
  to_csv --index=False example_results_to.csv
\end{lstlisting}

\begin{table*}
    \centering
    \caption{
        Tests-overview summarizing the results of our running example (only flaky tests).
    }{
        Omitted columns: Error\_sameOrder, Skipped\_sameOrder, Error\_randomOrder, Skipped\_randomOrder (all zero)
    }
    \label{tab:tests_overview}

\begin{tabular}{llllrrrrr}
\toprule
    Project\_Name          & \multilinecell{Project\_URL\\Project\_Hash\\Test\_filename\\Test\_classname\\Test\_param.} & Test\_name                                 & flaky?              & \multilinecellr{Passed\_\\same\\Order} & \multilinecellr{Failed\_\\same\\Order} & \multilinecellr{Passed\_\\random\\Order} & \multilinecellr{Failed\_\\random\\Order} \\
\midrule
  avwx-engine              & ...                                                                                                      & test\_fetch                                & non-order-dependent & 9                                      & 11                                     & 11                                       & 9 \\
flapy\_example             & ...                                                                                                      & test\_network\_remote\_connection\_failure & order-dependent     & 20                                     & 0                                      & 18                                       & 2 \\
flapy\_example             & ...                                                                                                      & test\_numpy\_random                        & non-order-dependent & 10                                     & 10                                     & 13                                       & 7 \\
flapy\_example             & ...                                                                                                      & test\_random                               & non-order-dependent & 15                                     & 5                                      & 8                                        & 12 \\
flapy\_example             & ...                                                                                                      & test\_time                                 & non-order-dependent & 12                                     & 8                                      & 9                                        & 11 \\
flapy\_example             & ...                                                                                                      & test\_unordered\_collections               & non-order-dependent & 9                                      & 11                                     & 11                                       & 9 \\
flapy\_example             & ...                                                                                                      & test\_victim                               & order-dependent     & 20                                     & 0                                      & 14                                       & 6 \\
      jgutils              & ...                                                                                                      & test\_numeric\_intermixed                  & non-order-dependent & 11                                     & 9                                      & 8                                        & 12 \\
      jgutils              & ...                                                                                                      & test\_file\_creation2                      & order-dependent     & 20                                     & 0                                      & 11                                       & 9 \\
\bottomrule
\end{tabular}

\end{table*}

Note that the directory specified after \texttt{--path} needs to be accessible from the current working directory since only the current working directory is mounted to the container that is started in the background.
The generated \textit{tests-overview} is saved to \texttt{./example\_results\_to.csv}.
\cref{tab:tests_overview} depicts its content.
Each row in the tests-overview corresponds to a test found and executed by pytest.
A test is uniquely identified by its name (Test\_name), the name of its class (Test\_classname), as well as possibly existing parametrizations (Test\_parametrization).
The eighth column (flaky?) contains FlaPy's verdict about if and how a test is flaky.
It can take four different values:

\begin{itemize}
    \item \textbf{not flaky}: no flaky behavior
        occurred, neither in same-order, nor in random-order runs.
    \item \textbf{non-order-dependent (NOD)}: flaky in same-order runs.
    \item \textbf{order-dependent (OD)}: flaky behavior in random-order runs, but not in same-order runs.
    \item \textbf{infrastructure}: flaky behavior, but only between iterations, never within.
        This suggests that a global issue occurred in some iterations that caused all runs of this test to fail.
\end{itemize}

The columns to the right of \enquote*{flaky?} show the number of times each verdict was reached in same- and random-order
(passed, failed, errored, or skipped).
We consider a test to be flaky, iff it yielded at least one passing and one failing or erroring run.
A test that alternates between failing and erroring runs is therefore not considered as flaky, since both outcomes would cause a build to fail and thus would not lead to the same developer experience as flakiness would cause.

Executing FlaPy on the three projects from our running example reveals nine flaky tests: six NOD and three OD.
Note that tests marked as OD might actually be NOD, if the flakiness never unveiled itself in the same-order runs, as happened in the case of \enquote*{\texttt{test\_network\_remote\_connection\_failure}}.
A recent study~\cite{lam2020understanding} showed that many NOD tests have very different failure rates for different orders.

\section{Evidence of Effectiveness}%

To demonstrate FlaPy's capabilities, we replicate our study on flaky tests in Python~\cite{gruber2021empirical}:
We randomly sample \num{30000} projects from PyPI~\cite{pypi} using FlaPy's \texttt{sample} command:
\begin{lstlisting}[language=bash,backgroundcolor=\color{white},numbers=none]
./flapy.sh sample --sample-size 30000 > flapy_input.csv
\end{lstlisting}
The script randomly samples projects from PyPI, searches for GitHub URLs, and removes duplicates.
Furthermore, it identifies PyPI tags for dependency installation wherever available an tries to match them with possibly existing Git tags to use the latter as the project hash.
If no matching Git tag is found, the field remains empty and the latest commit is used.
In our case, we retrieved a set of \num{26451} cloneable projects, whose test suites we run \num{100} times in default- and \num{100} times in random order.
Conducting the test executions on a SLURM cluster consisting of \num{58} machines, which allows us to run up to \num{700} iterations in parallel, %
takes about 3 days and generates 5~GB of compressed data.
In comparison, the total (i.e., sequential) runtime is about \num{20500} hours (\num{854} days).

Among the \num{10924} projects that contain at least one test, \num{396} also contain at least one flaky test.
In total, we find \num{2460} flaky tests in the roughly \num{1.5} million tests that were executed.
\num{1878} of them are OD, \num{543} are NOD, and \num{39} are flaky due to infrastructure.
These numbers confirm prior results~\cite{gruber2021empirical}, except for less infrastructure flakiness.
All data is accessible online~\cite{gruber2023flapyDataset}.
We hope that this new dataset will continue serving researchers, as our first one already has~\cite{wang2022ipflakies, wei2022preempting, ahmad2022identifying}.

\section{Extendability}%

FlaPy can be easily extended to implement new features, such as measuring further runtime metrics or manipulating the test execution to provoke flakiness. To do so, the \texttt{run} command offers the optional argument \texttt{--core-args CORE\_ARGS} to forward arguments directly to the core test execution procedure, which is the script \texttt{flapy/run\_tests.py}.
To extend the core, one can therefore easily add command line options to this script and activate them via the \texttt{CORE\_ARGS}.
Note that since the arguments will be unpacked later, they should be passed as one string (e.g.,~\texttt{--core-args "--foo --bar"}).
This could also be used to support projects written in languages other than Python.
One option that is currently available is \texttt{--random-order-seed}, which sets the seed for the test case shuffling, in case of random order runs.
After the core script has been modified, the FlaPy image needs to be rebuilt, which can be done via
\begin{lstlisting}[language=bash,backgroundcolor=\color{white},numbers=none]
docker build -t my_flapy -f Dockerfile .
\end{lstlisting}

This image can be used together with all existing scripts by changing the \texttt{FLAPY\_DOCKER\_IMAGE} variable in \texttt{setup\_docker\_command.sh} to \texttt{"localhost/my\_flapy"}.
When using this image in a SLURM execution, it needs to be pushed to a repository from which the cluster nodes can pull.

\section{Related work}%
\label{sec:related_work}

The iDFlakies framework~\cite{lam2019idflakies} is similar to FlaPy, as it also conducts repeated containerized test executions based on a set of projects that can be specified via their Git-URL and -Hash.
Unlike FlaPy, however, iDFlakies targets Java projects that are built using Maven or Gradle, and does not provide native support for parallelizing test executions.

Other rerun-based flakiness detection tools have been created by practitioners in the form of plugins for popular test execution frameworks~\cite{pythonRerunPlugin,mavenRerunPlugin,wendelin2020introducing,bazelAttributeFlaky}.
However, these approaches are not targeting researchers aiming to conduct studies and evaluations, but developers, aiming to detect flaky tests within their own projects. Therefore, they have weaker requirements in regard to isolation (since the code is known) and parallelization.
Aiming at the same goal, researchers proposed rerun-free flakiness detection approaches~\cite{bell2018deflaker,king2018towards,pinto2020what,kowalczyk2020modeling,alshammari2021flakeflagger,verdecchia2021know,parry2022evaluating,qin2022peeler,fatima2022flakify}.
FlaPy, on the other hand, targets researchers, who aim to conduct studies and evaluate their methods on large and diverse sets of sample projects.

\section{Conclusions}%
\label{sec:conclusions}

FlaPy is an easy to use tool to mine flaky tests by repeatedly executing the test suites of a given set of Python projects.
It is designed to address the challenges that come with this task:
By ensuring that the individual test runs are isolated, it simulates real-world CI conditions and ensures accurate results.
To promote diversity among the sample projects, FlaPy supports different forms of dependency management.
Furthermore, it provides means to parallelize the test executions across a SLURM cluster.
Lastly, FlaPy offers an easy-to-use interface to accumulate individual test outcomes into one concise table.
FlaPy's source code is available at \url{https://github.com/se2p/flapy} under the GNU LGPL license.
Contributions, feedback, and questions are most welcome.

\balance

\bibliography{main.bbl}

\noindent
All online resources accessed on Friday, 9th of February 2023.

\end{document}